\documentclass{ar-1col-S2O}
\usepackage[numbers]{natbib}
\usepackage{url}
\usepackage{unicode}
\usepackage{makecell}
\usepackage{amssymb}
\usepackage{xcolor}
\usepackage{textcomp}
\usepackage{siunitx}
\usepackage{caption}
\usepackage{lipsum}
\usepackage{lineno}

\usepackage{graphicx}
\graphicspath{{figures/}}

\newcommand{\micrometer}{$\mathrm{\mu}$m }
\newcommand{\microsecond}{$\mathrm{\mu}$s }

\makeatletter
\def\journalinfo{{\jinfofont
  Posted with permission from the \textit{Annual Review of Condensed Matter Physics}, Volume~18 \textcopyright\ 2027 by Annual Reviews, \url{https://www.annualreviews.org}.\endgraf
  \vskip6\p@
  Published version:\endgraf
  \url{https://doi.org/10.1146/annurev-conmatphys-031424-024231}\endgraf}}
\makeatother

\begin{document}

\markboth{Sorichetti et al.}{Beads, springs and fields: particle-based \emph{vs} continuum models in cell biophysics}

\title{Beads, springs and fields: particle-based \emph{vs} continuum models in cell biophysics}

\author{Valerio Sorichetti,$^{1,2*}$ Juraj M\'ajek,$^{1*}$ Ivan Palaia,$^{1,3*}$ Fernanda P\'erez-Verdugo,$^{1*}$ Christian Vanhille-Campos,$^{1,4}$ Edouard Hannezo,$^1$ and An{\dj}ela \v{S}ari\'c$^{1\dag}$
\affil{$^1$Institute of Science and Technology Austria, 3400 Klosterneuburg, Austria}
\affil{$^2$Institute of Molecular Biotechnology of the Austrian Academy of Sciences (IMBA), Vienna BioCenter (VBC), 1030 Vienna, Austria}
\affil{$^3$Department of Physics, King’s College London, London WC2R 2LS, United Kingdom}
\affil{$^4$Laboratoire Jean Perrin, CNRS/Sorbonne Université, 75005 Paris, France}
\affil{* These authors contributed equally}
\affil{\dag email: andela.saric@ist.ac.at}
}

\begin{abstract} 
Quantitative modeling has become an essential tool in modern biology, driven by advances in both experimental techniques and theoretical frameworks. Powerful high-resolution techniques now provide detailed datasets spanning molecular to tissue scales, allowing us to visualize cellular structures with unprecedented detail. In parallel, developments in soft and active matter physics have established a robust theoretical basis for describing biological systems. In this context, two main modeling paradigms have emerged: particle-based models, which explicitly represent discrete components and their interactions, and continuum models, which describe systems through spatially varying fields. We compare these approaches across biological scales, highlighting their respective strengths, limitations, and domains of applicability. To keep our discussion biologically relevant, we focus on five systems of fundamental importance: the cytoskeleton, membranes, chromatin, biomolecular condensates and tissues. With this Review, we thus aim to provide a framework for both theorists and experimentalists to select appropriate modeling strategies, and highlight future directions in biophysical modeling.
\end{abstract}

\begin{keywords}
biophysical modeling, cell mechanics, cytoskeleton, membranes, biomolecular condensates, chromatin, biological tissues \end{keywords}
\maketitle

\tableofcontents

\section{INTRODUCTION}

\subsection{Modeling in biophysics}
\label{sec:intro}

In the past few decades, quantitative modeling has gained primary importance in cell and tissue biology~\cite{poon2006soft,ramaswamy2010mechanics,marchetti2013hydrodynamics,doi2013soft,Prost2015,hafner2019minimal}. The shift to a more quantitative approach has been driven both by technological and theoretical developments. On the technological side, \textbf{breakthroughs in experimental techniques} have given access to new types of data, often of unprecedented quality and in vastly greater quantity compared to what has been accessible in the past. Super-resolution imaging methods~\cite{sigal2018visualizing} and cryo-electron microscopy~\cite{nogales2024bridging}, together with single-molecule tracking techniques~\cite{kusumi2014tracking}, have enabled researchers to visualize cellular structures and their dynamics with unprecedented detail down to the molecular level. In parallel, methods such as Hi-C~\cite{lieberman2009comprehensive} and DNA tracing~\cite{bintu2018super} have allowed the study of the 3D organization of the genome. Furthermore, the refinement of force-measuring techniques such as atomic~\cite{krieg2019atomic} and traction~\cite{style2014traction} force microscopy, magnetic and optical tweezers~\cite{catala2022exploring}, and laser ablation~\cite{zulueta2015laser} has allowed scientists to measure forces in living cells and tissues~\cite{roca2017quantifying}. At the multicellular scale, light-sheet microscopy has enabled real-time three-dimensional visualization
of living tissues~\cite{girkin2018light}, and organoids made it possible to study realistic 3D tissues \emph{in vitro}~\cite{rossi2018progress}.

Equally important as these technological developments have been those in the theory of soft and active matter, which have contributed in laying a solid theoretical foundation for the quantitative understanding of biological phenomena at the cell and tissue scales. \textbf{Soft matter physics}~\cite{poon2006soft,doi2013soft} has provided the conceptual and methodological tools to describe polymers (such as DNA and proteins), membranes, biological condensates and ensembles of cells. \textbf{Active matter physics}~\cite{ramaswamy2010mechanics,marchetti2013hydrodynamics} has additionally laid out a framework to understand systems ---such as cells--- which convert energy into motion, force generation and self-organization. Driven by these technological and theoretical advancements, biophysical modeling has established itself as a fundamental tool to help us interpret experimental data, formulate new predictions and uncover the general physical principles behind living systems.

In this context, two modeling paradigms have been particularly successful: \textbf{particle-based models}, which retain the discreteness of the building blocks and interactions present in the system, and \textbf{continuum models}, which describe the system \emph{via} continuous fields representing the average behavior of the microscopic components and their interactions. We aim to offer an overview and comparison of the most used particle-based and continuum models in cell biophysics, highlighting their respective strengths and weaknesses. 

This Review thus has a triple purpose: first, to guide theorists in choosing the most suitable modeling approach to address their biological systems of interest. Second, to provide experimentalists with an overview of the most prevalent biophysical models, and to highlight which ones might be best suited to complement their experiments. Third, to emphasize the next frontiers of biophysical modeling in the coming years.

\subsection{Particle-based and continuum models}

\begin{figure}[t]
\includegraphics[width=0.9\textwidth]{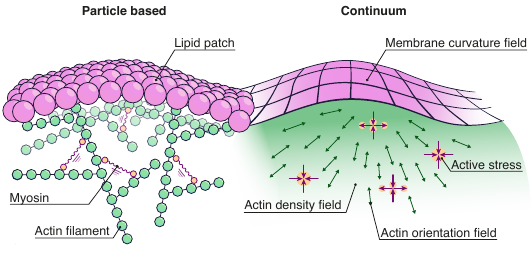}
\caption{Schematic illustration of particle-based and continuum models for the cell membrane and the actin cortex. In a particle-based model (left), lipid patches, filaments and motors can be represented as a collection of beads whose mutual interactions regulate their mechanical properties. In a continuum model (right), the spatially-varying properties of the system are represented by continuous fields. Figure created by C. Flandoli based on an original concept by V. Sorichetti.}
\label{fig:part1_summary}
\end{figure}

\textbf{Particle-based models} describe the system as a collection of discrete objects (atoms, molecules, molecular aggregates or cells) interacting \emph{via} simple physics-based rules (Fig.~\ref{fig:part1_summary}, left). These models fall into the wider class of \textbf{agent-based models}, where the modeled entities and their interactions can be more generally defined~\cite{grimm2005pattern}. Typically, particle-based models are simulated employing numerical techniques such as \textbf{molecular dynamics} or \textbf{Monte Carlo} methods~\cite{allen2017computer,frenkel2023understanding}.%
\begin{marginnote}[]\entry{Molecular dynamics}{Computational method that simulates the dynamics of particles by solving Newton's equations of motion.}\entry{Monte Carlo}{Computational method that samples equilibrium configurations of a system using  stochastic algorithms.}\end{marginnote}%
\textbf{Continuum models} describe the system as a medium whose properties (such as the concentrations of molecules, their velocities and orientations) vary continuously in space and can be described by mathematical functions called \textbf{fields}~\cite{poon2006soft,doi2013soft} (Fig.~\ref{fig:part1_summary}, right). The equations describing these fields can be solved either analytically or numerically. Sometimes, a discretization of the fields and/or of the equations is adopted to allow the numerical solution of the model~\cite{reddy2010finite}, but this is done only for computational convenience and bears no direct connection with the discreteness of the underlying system, in contrast with particle-based models.\begin{marginnote}[]
\entry{Field}{Function describing the variation of a physical observable over space and time.}\end{marginnote}

\subsection{Coarse-graining: bottom-up and top-down}

One could think that a possible approach to modeling biological systems would be that of describing them down to the atomistic level. However, cellular systems contain a staggering number of atoms and molecules, ranging from order $10^6$ atoms in a 1-\micrometer actin filament to up to $10^{15}$ in a whole cell. This enormous number of \textbf{degrees of freedom} makes a full atomistic description impractical beyond very small systems.%
\begin{marginnote}[]\entry{Degrees of freedom}{Independent variables that specify the state of a physical system.}\end{marginnote}%
To give a more concrete measure of the complexity involved, even simulating something as simple as a  $65$-nm bacteriophage for a few \microsecond requires advanced and highly resource-intensive methods~\cite{coshic2024structure}. Moreover, even if atomistic simulations of an entire cell were feasible, it is unclear how useful they would be for interpreting experiments or making predictions, since the complexity of the model would then match that of the real system~\cite{goldenfeld1999simple,phillips2012physical}.

A powerful approach to develop biophysical models that are both tractable and interpretable is \textbf{coarse-graining}~\cite{poon2006soft,peter2009multiscale,hafner2019minimal}, where the system is replaced by a reduced representation with fewer degrees of freedom, enabling analytical or computational tractability. The complexity of the real system is then encoded in a few effective parameters.

In coarse-grained \textbf{particle-based models}, discrete units representing groups of atoms or molecules interact via effective potentials~\cite{likos2001effective}. These potentials, which encode parameters such as the strength and range of the inter-particle interactions, are often designed to capture universal properties of the system. For example, many structural and dynamical features of proteins, DNA, and cytoskeletal filaments can be qualitatively reproduced by representing them as beads connected by springs, \emph{i.e.} \textbf{bead-spring polymers}~\cite{hafner2019minimal}.%
\begin{marginnote}[]\entry{Bead-spring model}{Particle-based model in which molecules are represented as collection of beads connected by elastic springs (bonds).}\end{marginnote}%

\textbf{Continuum models} achieve an even higher level of coarse-graining than particle-based models, sacrificing microscopic detail for analytical or computational simplicity and greater interpretability. For example, a coarse-grained particle-based model of the cell cytoskeleton would involve simulating tens of thousands of individual particles representing filament segments, motors and cross-linkers (Sec.~\ref{sec:cytoskeleton_pb}). A continuum approach like active gel theory (Sec.~\ref{sec:cytoskeleton_cm}), on the other hand, can describe the large-scale behavior of the system with only a handful of equations.

Both particle-based and continuum coarse-grained models can be constructed following two approaches: bottom-up and top-down. In the \textbf{bottom-up} approach, model ingredients and parameters are derived from more detailed (\emph{e.g.} atomistic) models. The accuracy of the coarse-graining procedure thus plays a key role in this model building approach. In contrast, the \textbf{top-down} approach starts from general physical principles, such as symmetries and conservation laws, to decide which ingredients to include. Parameters are then chosen to reproduce the desired behavior. Top-down models typically aim to explain qualitative trends in experiments and to identify different regimes in parameter space.

\section{MODELING BIOLOGICAL SYSTEMS}
\label{sec:part2}

In this Section, we discuss and compare some of the successes of continuum and particle-based approaches in cell and tissue biophysics. While describing the merits of both approaches, we highlight how they have been complementary in addressing specific questions by discussing selected case studies.

\subsection{Cytoskeleton}
\label{sec:cytoskeleton}

The mechanics of the cytoskeleton results from the dynamics of filaments such as actin and microtubules, motors and cross-linkers, as well as their interaction with the membrane and the rest of the cytoplasm~\cite{howard2002mechanics}. Filaments can nucleate and polymerize from specific cytoplasmic proteins, are stochastically cross-linked, and are displaced by molecular motors. Together, they form dynamic cellular structures such as the spindle and the actomyosin cortex, which provide a localized scaffold capable of generating active forces to reshape the cell. 

\begin{figure}
    \centering
    \includegraphics[width=0.99\linewidth]{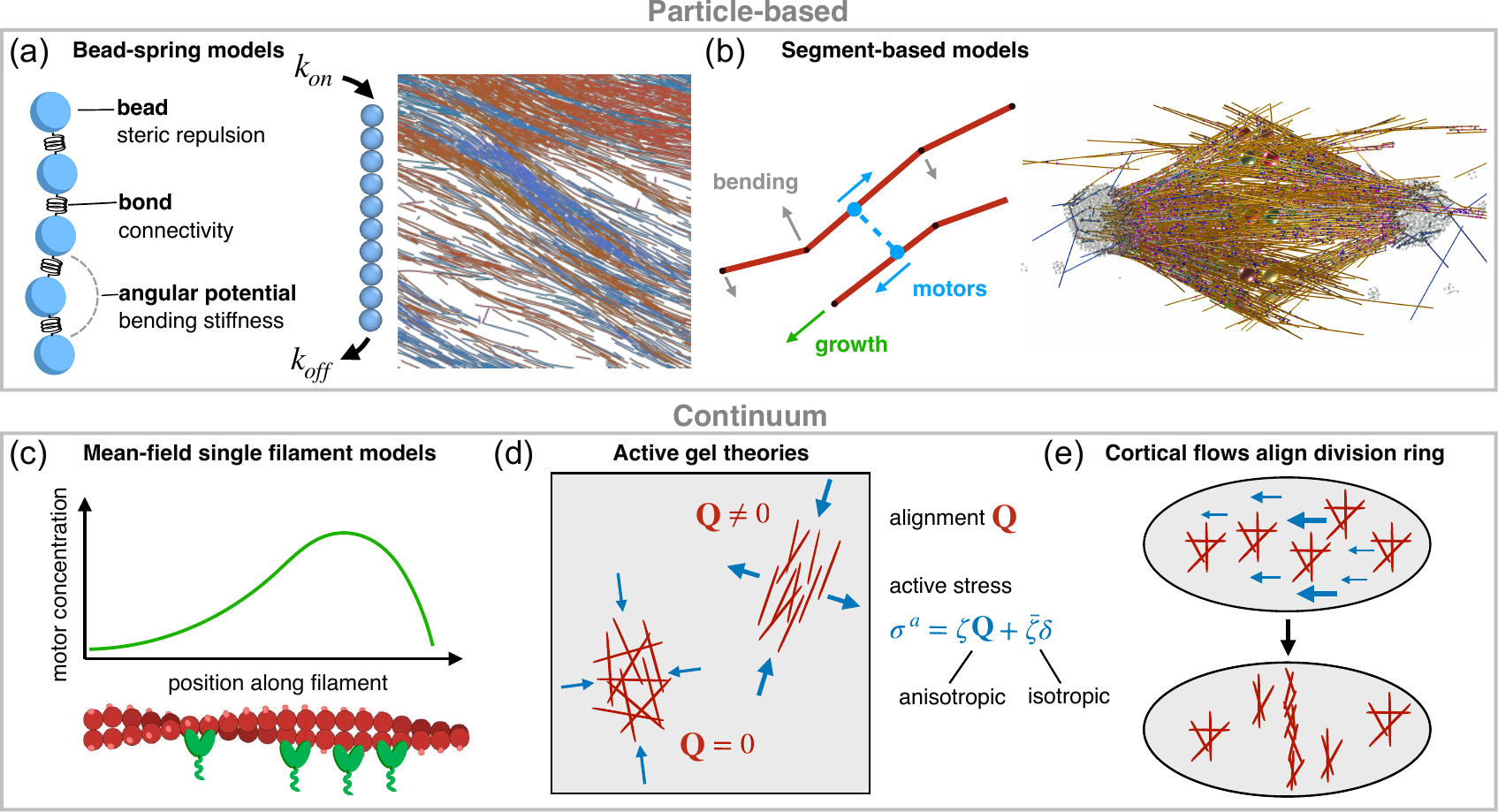}
    \caption{\textbf{Models for the cytoskeleton.} \textbf{(a)}
    Left: In a bead-spring description, filaments are represented as beads connected by elastic bonds. These models can be used to resolve filament growth and shrinkage during treadmilling at the monomer level. Right: Collective nematic alignment of treadmilling filaments, simulated with the model of Reference~\citenum{vanhille2024self} (color coding reflects filament orientation).
    \textbf{(b)} Left: In the segment-based description, filaments are represented as a collection of line segments connected by hinges~\cite{Nedelec2007}. Filaments can grow or shrink and motors can walk along them. Right: Model of the mitotic spindle~\cite{motta2024cell} simulated in Cytosim~\cite{Nedelec2007}. Microtubules are colored according to their nucleation pathway, with most of them being nucleated through the augmin pathway (yellow). Image courtesy of François Nédélec. \textbf{(c)} In single-filament continuum models, the average density of motors (vertical axis) along the filament length (horizontal axis) is derived by comparing binding/unbinding and treadmilling rates, and used to compute the resulting force~\cite{Lenz2020}. Sketch adapted from CNX OpenStax/Wikimedia Commons~\cite{actinsketch} (CC BY 4.0). \textbf{(d)} In active-gel models, the cytoskeleton is a continuum field with a polar or nematic orientation, while motor contractility (blue arrows) is modeled by an orientation-dependent active stress term~\cite{Prost2015}. \textbf{(e)} Active gel theory was used to describe the formation of the cytokinetic furrow in \emph{C. elegans}~\cite{Reymann2016}, as described in Sec.~\ref{sec:cytoskeleton_cs}.}
    \label{fig:cytoskel}
\end{figure}

\subsubsection{Particle-based} 
\label{sec:cytoskeleton_pb}
Particle-based models explicitly resolve individual objects within the cytoskeleton. Filaments, monomers and motors can be modeled as assemblies of either spherical beads or cylindrical segments. This enables modifying their individual characteristics to pinpoint how they influence the macroscopic properties of the cytoskeleton, such as connectivity and contractility. For instance, this approach has helped understand how end-tracking cross-linkers and motors can generate contractile or extensile forces~\cite{MendesPinto2012,Belmonte2017,Wollrab2019}, and explained why force generation is maximized for intermediate filament lengths~\cite{Chugh2017}.

Particle-based models can describe filaments with varying levels of detail, either resolving each monomer with a \textbf{bead-spring model} \cite{duman2018collective,vanhille2024self} (Fig.~\ref{fig:cytoskel}a) or representing a semiflexible filament as a collection of line segments connected by hinges (Fig.~\ref{fig:cytoskel}b). The latter \textbf{segment-based models} involve fewer degrees of freedom, and are thus more computationally efficient than bead-spring ones. The segment-based approach has been employed by most software developed for the cytoskeleton specifically, such as Cytosim~\cite{Nedelec2007}, MEDYAN~\cite{Popov2016,Ni2021}, AFINES~\cite{Freedman2017} and aLENS~\cite{Yan2022}. However, to achieve this speed-up, compromises often need to be made in how accurately these models resolve inter-filament steric repulsion or cross-linker stiffness. Bead-spring models, while slower, naturally account for steric effects and can more easily address situations where entanglement between filaments is crucial~\cite{DarPalaia}. Additionally, some processes, such as treadmilling~\cite{erlenkamper2012impact,vanhille2024self} are more naturally simulated using models where single monomers are explicitly resolved.

\subsubsection{Continuum} 
\label{sec:cytoskeleton_cm}
Several continuum models represent a single filament as a line segment, and model the distribution of motors, cross-linkers and mechanical tension across its length (Fig.~\ref{fig:cytoskel}c).
They usually treat interactions with the neighboring network at the \textbf{mean-field} level.%
\begin{marginnote}[]
\entry{Mean-field approximation}{The replacement of an object's local, instantaneous interactions with neighbors by an average interaction based on its surrounding environment.}
\end{marginnote}%
This amounts to assuming that understanding the average environment of the single filament is sufficient to explain network properties. These models were instrumental to propose filament buckling~\cite{Lenz2012} and cross-linker or motor organization~\cite{Kruse2000,Belmonte2017,Lenz2020} as possible causes for the emergence of contractility in actomyosin.

A prominent role is played by \textbf{active gel theory}~\cite{Prost2015}, where cytoskeletal networks at the cell scale are coarse-grained to a filament density field, endowed with orientation and  active stresses resulting from the action of motors (Fig.~\ref{fig:cytoskel}d)~\cite{Kruse2005}. These stresses are added to those of a passive viscoelastic medium, resulting in flows (Fig.~\ref{fig:cytoskel}e) which affect the density and orientation fields. Active gel models are thus ideally suited to explore large-scale systems such as the actin cortex, and have helped understand cytokinesis~\cite{Salbreux2009,Turlier2014,Reymann2016} (Sec.~\ref{sec:cytoskeleton_cs}) and spindle organization~\cite{Brugues2014,Middelkoop2024}. These models can also be adapted to include feedback loops and mechanochemical couplings within the cortex~\cite{Bois2011,Nishikawa2017}, and are additionally employed to describe tissue mechanics (Sec.~\ref{sec:multicell_cm}). 

\subsubsection{Comparison} 
Continuum models often describe the cytoskeleton in a top-down manner, agnostic to exact microscopic details. They can also be easily coupled with \textbf{reaction-diffusion models} for upstream signaling factors~\cite{Nishikawa2017,drozdowski2023optogenetic}.%
\begin{marginnote}[]\entry{Reaction-diffusion model}{Mathematical framework describing how chemical species spread in space and time while undergoing chemical reactions.}\end{marginnote}%
However, integrating them with models of other cellular structures, such as membranes, can be harder than for bead-spring particle-based models, where all structures are built of beads and are thus inherently compatible~\cite{Sciortino2025}. Whereas particle-based models naturally account for the detailed local structure of the cytoskeleton, continuum ones often neglect correlations between filaments, resulting \emph{e.g.} from the local cortex architecture~\cite{Chugh2017} and from \textbf{entanglements}~\cite{DarPalaia}.%
\begin{marginnote}[]\entry{Entanglements}{Topological constraints emerging in dense systems where polymers cannot cross, restricting motion and producing slow, collective relaxation dynamics.}\end{marginnote}%

\subsubsection{Case study: Cytokinesis}
\label{sec:cytoskeleton_cs}

Particle-based and continuum models have both been used, in complementary ways, to study cytokinesis, the final stage of cell division in which a contractile ring of aligned actin filaments forms at the cell equator and constricts to divide the cell. Two key questions arise in this context: how this ring forms, and how it subsequently constricts.

Ref.~\citenum{Descovich2018} used a segment-based particle model to model the ring. This approach allowed the authors to quantify how constriction speed depends on the concentration of myosin motors and cross-linkers. While the model reproduced experimentally observed trends for varying cross-linker concentration, it failed to capture the qualitative dependence on myosin concentration. The authors suggested that this discrepancy arises because myosin also influences the speed of formation of the ring, not only constriction speed itself.

This missing ingredient had been addressed in Ref.~\citenum{Reymann2016}, which studied ring formation using an active-gel continuum model. The authors showed that the ring is formed by large-scale cortical flows that compress and align actin filaments at the cell equator. Because these flows extend across the entire cortex, capturing this mechanism requires a cell-scale description. Ref.~\citenum{Descovich2018} modeled only the ring itself, as simulating the full cortex with a particle-based model was computationally prohibitive. As a result, the model of Ref.~\citenum{Descovich2018} could not account for the full effect of myosin on ring formation dynamics.

These two studies illustrate how particle-based and continuum models can answer different questions about the same system. The particle-based approach can resolve how biological processes depend on the concentration of specific components, such as motors and cross-linkers. In contrast, continuum models are better suited to describe cell-wide cytoskeletal dynamics.

\subsection{Membranes}
\label{sec:membranes}

\begin{figure}
    \centering
    \includegraphics[width=0.99\linewidth]{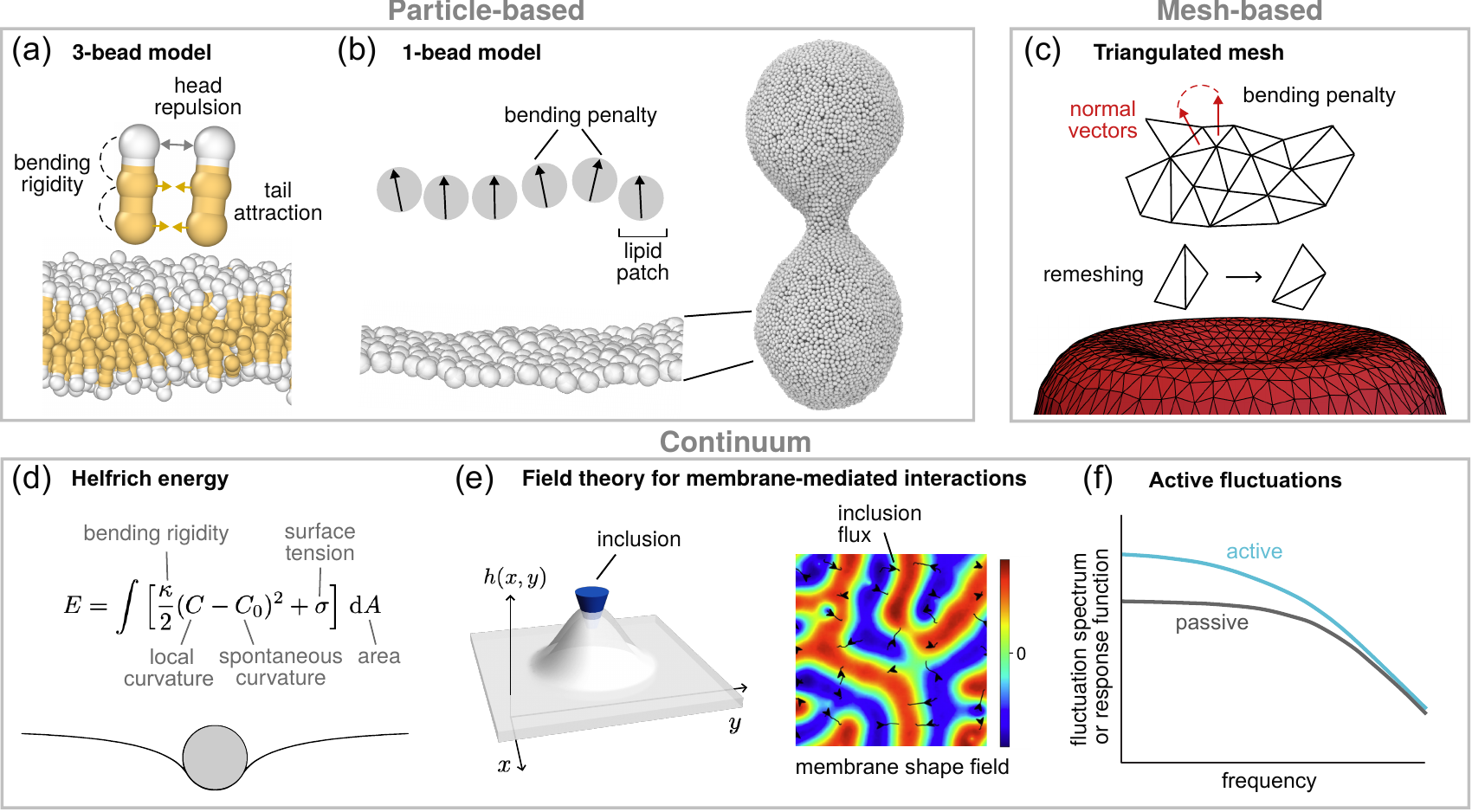}
    \caption{\textbf{Models for biological membranes.} \textbf{(a)} Lipids modeled as 3-bead polymers. Heads and tails are beads of different types, interacting with each other (arrows). A bending potential (dashed lines) straightens the lipid~\cite{amaral2025stability}. \textbf{(b)} Left: In fluid one-bead models, a bending penalty opposes the mutual misalignment of oriented beads (arrows), encoding rigidity. Right: A fluid vesicle constricted by an external force, simulated with the model of Reference~\citenum{yuan2010one}. \textbf{(c)} A triangulated mesh representing a surface. Fluidity is ensured by edge flipping and rigidity by a penalty for misaligned normal vectors (arrows). \textbf{(d)} The Helfrich theory treats the membrane as a continuum and, in its simplest formulation, assigns it a bending energy, depending on local curvature, and a surface energy, proportional to its area. This can explain membrane deformations around adhered objects~\cite{Deserno2004}. Panel adapted with permission from Reference~\citenum{Deserno2004}; copyright 2004 American Physical Society. \textbf{(e)} Field theories for membrane inclusions model adhered proteins as domains that constrain the local membrane properties (\textit{e.g.} curvature). Inclusions, interacting \emph{via} a field representing membrane shape, move to and cluster in regions of preferred geometry~\cite{Bitbol2019,Zakine2018}. Panel adapted with permission from Reference~\citenum{Zakine2018}; copyright 2018 American Physical Society. \textbf{(f)} Often complemented with a Langevin dynamics, the Helfrich theory can predict fluctuation spectra, which are quantitatively different in active and passive membranes \cite{Turlier2016,Sciortino2025}.
    }
    \label{fig:membrane}
\end{figure}

Biological membranes are thin fluid sheets that resist stretching, but can bend. The high mobility of the constituent lipid molecules allows them to rapidly exchange neighbors, making membranes fluid and able to reshape~\cite{Helfrich1973}. By producing or uptaking new lipids, cells can additionally change the membrane area dynamically. Morphological changes are a major focus of membrane models: the choice of which model to employ will typically depend on the length scales of interest and on the origin of the deformation forces (\emph{e.g.} interaction with cytoskeletal filaments or other proteins).

\subsubsection{Particle-based models} 

A common approach to coarse-graining membranes is to represent individual lipids using a bead-spring model, \emph{e.g.} as \textbf{3-bead} units where one bead represents the lipid head and two represent the tail. Hydrophobic interactions are then modeled as an effective attraction between the tails (Fig.~\ref{fig:membrane}a). These models allow the study of large scale deformations and scission phenomena, such as budding or neck constriction by molecular machinery, and yet retain the lipid discreteness and control over the composition of the membrane~\cite{cooke2005solvent,cooke2005tunable,reynwar2007aggregation}. In Ref.~\citenum{amaral2025stability}, for instance, models with 3-bead lipids forming a bilayer and 6-bead lipids forming a monolayer were used to compare the mechanics of these two membrane structures.

\textbf{One-bead-thick} models of membranes represent a whole lipid patch as a single bead, which can freely diffuse in the membrane plane, ensuring fluidity~\cite{yuan2010one} (Fig.~\ref{fig:membrane}b). These types of models have proven useful for complex large-scale reshaping processes in the cell, such as budding~\cite{Jiang2022}, protrusion formation~\cite{Paraschiv2021,Wang2024}, cell division~\cite{TarrasonRisa2020,Harker-Kirschneck2022} or osmotic shocks~\cite{vanhille2021}. These models can also include hydrodynamic effects, which can be important to capture \emph{e.g.} domain growth dynamics~\cite{laradji2004dynamics} and the role of cytoplasmic viscosity on membrane fluctuations~\cite{Sadeghi2018}.

\subsubsection{Mesh-based models}

Mesh-based models are situated at the boundary between particle-based and continuum models~\cite{munoz-basagoiti2025tutorial}. In these models, the membrane is discretized as a network of vertices connected by edges, which define small flat surface portions, usually resulting in a \textbf{triangulated mesh}~\cite{gompper1997network,noguchi2009membrane,kumar2022review} (Fig.~\ref{fig:membrane}c). Albeit similar in spirit to particles in one-bead-thick models, vertices in mesh models do not necessarily represent lipid patches of constant area, but are rather the discretization of a continuum surface whose energy is typically described by the Helfrich theory (Sec.~\ref{sec:membranes_cm}). In mesh-based models, membrane fluidity is incorporated through edge flipping (\emph{remeshing}). Vertex positions can be evolved using molecular dynamics, Monte Carlo, or hybrid techniques~\cite{siggel2022trimem,matoz2023pymembrane,munoz-basagoiti2025tutorial}, or simply by energy minimization, \emph{e.g.} with Surface Evolver~\cite{brakke1992surface}. Mesh-based models can also be readily coupled to particle-based models, as in Ref.~\citenum{Ni2021}, where they were combined with a segment-based description of the cytoskeleton.

\subsubsection{Continuum models} 
\label{sec:membranes_cm}
Continuum models describe membranes as 2D surfaces embedded in 3D space. At scales larger than a few lipids, the mechanics and energetics of the bilayer are assumed to be governed by its geometry, given material properties such as rigidity, spontaneous curvature, elasticity, and fluidity~\cite{seifert1997configurations,deserno2015fluid,lipowsky2022remodeling}. In this framework, geometry determines the energy, which in turn defines equilibrium configurations and fluctuation spectra through statistical mechanics.

The successful \textbf{Helfrich model}~\cite{Helfrich1973,Canham1970,Evans1974} describes the energy of a membrane element as the sum of a bending term, proportional to the squared curvature, and a tension term, proportional to the area  (Fig.~\ref{fig:membrane}d). The membrane is implicitly assumed to be infinitely fluid, thus deformable. Adhesion can also be introduced, and for simple configurations it is possible to include hydrodynamics~\cite{seifert1997configurations}. Besides serving as an immediate benchmark for most particle- or mesh-based models, the Helfrich model has made it possible to study how vesicles are deformed by external stresses~\cite{Seifert1991,Lipowsky1991,Seifert1995a,deserno2015fluid,lipowsky2022remodeling}, or how membranes can adhere to other surfaces~\cite{Seifert1990}, wrap particles~\cite{Deserno2004,Deserno2004a} and produce tubes and tethers by extrusion~\cite{smith2004pulling,Derenyi2002}. In the latter case, rigidity and tension have been related to the extrusion force and the resulting tube radius: this provided a way to measure mechanical properties from simple experiments~\cite{Evans1987,Cuvelier2005,Roux2013} or in simulations~\cite{Paraschiv2021,munoz-basagoiti2025tutorial}. By coarse-graining proteins that bind to the membrane to modify its preferred curvature as a field, one can reproduce complex membrane architectures observed in the cell~\cite{Hu2008,Sadhu2023}. Continuum models were also used to understand how rigid and deformable particles adhering to a membrane interact with each other \emph{via} the deformations they induce on it~\cite{Goulian1993,Kim1998,reynwar2011membrane,Schweitzer2015,Bitbol2019,Zakine2018,Fournier2022} (Fig.~\ref{fig:membrane}e and Sec.~\ref{sec:membrane_cs}).

Active processes arising \emph{e.g.} from the cytoskeleton~\cite{Maitra2014} or energy-consuming inclusions~\cite{Cagnetta2022} bring membranes out of equilibrium. In these \textbf{active membranes}, fluctuations and wave propagation differ fundamentally from their passive counterparts and do not obey the \textbf{fluctuation–dissipation relation} (Fig.~\ref{fig:membrane}f).%
\begin{marginnote}[]\entry{Fluctuation-dissipation relation}{Mathematical relation connecting a system’s spontaneous fluctuations at equilibrium to how it responds to small external perturbations.}\end{marginnote}%
This behavior has been observed experimentally and mostly explained by continuum models (for small deformations), particularly in studies of flickering red blood cells~\cite{Turlier2016} and vesicles containing cytoskeletal filaments~\cite{Sciortino2025}.

\subsubsection{Comparison}

Continuum models of membranes are best suited to study equilibrium shapes and deformations with simple geometry, such as tether pulling~\cite{smith2004pulling,Derenyi2002}. Their applicability, however, becomes limited when dealing with complex shapes or large-scale nonlinear deformations. Moreover, they cannot easily incorporate properties such as heterogeneous lipid composition and leaflet asymmetry. 
Mesh-based models can be more efficient to study complex large-scale deformations, and can to a certain extent model local heterogeneities. However, simulating topology changes such as membrane fission and fusion with these models is not trivial~\cite{tachikawa2017golgi}. Similarly, capturing extreme deformations, like the emergence of thin necks during particle wrapping~\cite{azadbakht2024nonadditivity}, requires very fine, computationally expensive meshes. Particle-based models can more easily capture large deformations and topology changes, but are limited to smaller system sizes. Constraints on vesicle volume are not trivial to implement in particle-based simulations and may require \emph{e.g.} the use of filler particles~\cite{vanhille2021}, whereas they are more straightforward in mesh-based and continuum models.

\subsubsection{Case study: Effective interactions between membrane proteins} 
\label{sec:membrane_cs}

Membrane inclusions such as transmembrane proteins can impose local constraints on membrane shape and ultimately result in effective interactions between inclusions. This phenomenon was initially pointed out in Ref.~\citenum{Goulian1993} within the small-deformation limit of the Helfrich model. In this framework, effective interactions arise between inclusions because these change both the curvature of the membrane and its fluctuations~\cite{Goulian1993,fournier1997comment}. Fluctuation-induced interactions are attractive~\cite{Goulian1993,golestanian1996fluctuation}; however, in typical biological conditions, these are expected to be overwhelmed by curvature-mediated forces, which for approximately flat membranes are predicted to be repulsive~\cite{reynwar2011membrane,Bitbol2019}. These predictions are however limited to the analytically tractable small-deformation regime, far from many biologically relevant situations.

Particle-based simulations were instrumental in overcoming this limitation: in Ref.~\citenum{reynwar2007aggregation} it was shown that particles that impose a strong local curvature on the membrane can qualitatively alter the shape of the effective interactions: an attractive regime appears at intermediate distances, ultimately inducing protein clustering and vesicle formation. This picture was later confirmed within continuum elasticity beyond the linear regime, combining mesh-based and analytical approaches in Ref.~\citenum{reynwar2011membrane}. A particularly well-studied example is that of BAR proteins, whose strongly curved shape requires mesh- or particle-based models able to resolve both high curvatures and large deformations (see \emph{e.g.} Ref.~\citenum{Ayton2009,simunovic2013linear}). 

Continuum theories thus established the physics of curvature-mediated interactions in the small-deformation regime, while particle- and mesh-based simulations revealed qualitatively new behavior at large deformations, including the emergence of attractive interactions and protein aggregation, thereby extending these ideas to more biologically relevant situations.

\subsection{DNA and chromatin}
\label{sec:chromatin}

In eukaryotes, the genome is compacted and organized by numerous specialized proteins, resulting in the chromatin fiber. Due to its polymer-like properties, chromatin ---like cytoskeletal filaments (Sec.~\ref{sec:cytoskeleton_pb})--- lends itself to being modeled as a bead-spring polymer. This particle-based approach has been invaluable to understand the small-scale organization of the genome and its dynamics. The larger-scale organization, such as that governing euchromatin and heterochromatin regions, can instead be more easily understood with the aid of continuum models.

\subsubsection{Particle-based models} 
\begin{figure}[t]
\includegraphics[width=\textwidth]{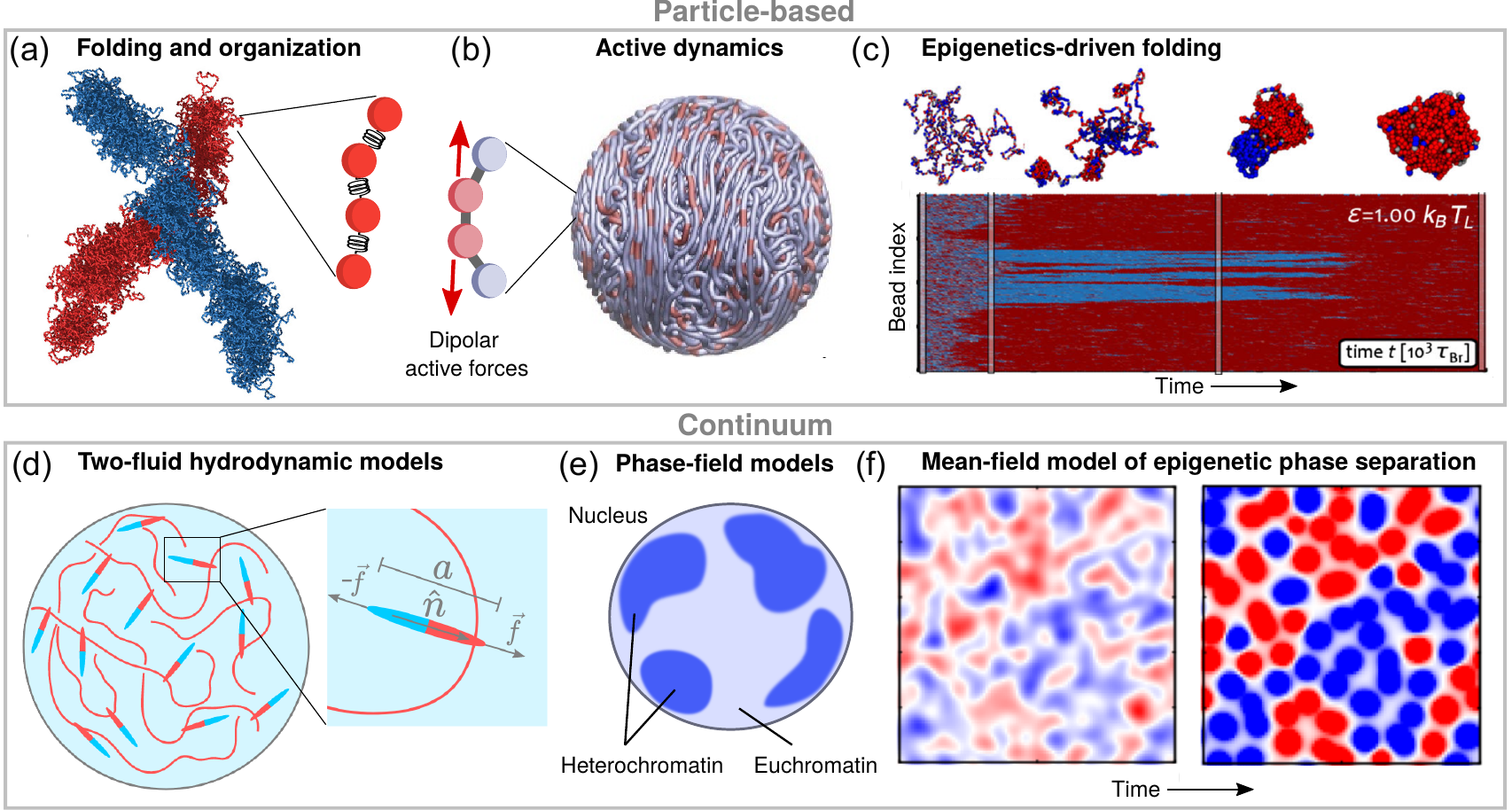}
\caption{\textbf{Models for chromatin.} \textbf{(a)} Bead-spring model of two mitotic sister chromatids (red and blue), connected by the centromere and individualized by the action of loop extruders, modeled as harmonic bonds that "walk" along the chain. Panel adapted from Reference~\citenum{goloborodko2016compaction} (CC BY 4.0). \textbf{(b)} Chromatin fiber with active extensile dipolar forces with red segments showing instantaneous dipole locations. Panel adapted with permission from Reference~\citenum{saintillan2018extensile}. \textbf{(c)} Chromatin model with epigenetic domains (red and blue colors). Starting from a randomly colored polymer, distinct epigenetic domains emerge over time (kymograph) due to chromatin-mediated interactions and stochastic switching, until a single epigenetic mark takes over. Panel adapted from Reference~\citenum{michieletto2016polymer} (CC BY 3.0).  \textbf{(d)} Schematic of a two-fluid chromatin model. Polar (orientation $\hat n$) motors of size $a$ exert equal and opposite forces $\pm\vec f$ on polymer and solvent. Panel adapted with permission from Reference~\citenum{eshghi2023activity}; copyright 2023 American Physical Society. \textbf{(e)} Schematic representation of a phase-field model of the nucleus, with euchromatin regions in light blue and heterochromatin in dark blue. \textbf{(f)} Microphase separation between epigenetic domains (red and blue colors) in a magnetic model of chromatin. Panel adapted with permission from Reference~\citenum{michieletto2019nonequilibrium}; copyright 2019 American Physical Society.}
\label{fig:chromatin}
\end{figure}

Particle-based models of chromatin allow for a straightforward comparison with experimental structural~\cite{lieberman2009comprehensive,mateos2009spatially,naumova_organization_2013,nuebler2018chromatin} and dynamical data~\cite{saintillan2018extensile,shi2018interphase,mahajan2022euchromatin,shin2024transcription}. Consequently, they have made a significant contribution to our understanding of the organization and reshaping mechanisms of eukaryotic chromosomes, both in mitosis~\cite{naumova_organization_2013,goloborodko2016chromosome,goloborodko2016compaction,hildebrand2024mitotic}, \emph{e.g.} explaining the origin of the compact cylindrical shape of mitotic chromosomes (Fig.~\ref{fig:chromatin}a and Sec.~\ref{sec:chromatin_cs}), and in interphase~\cite{rosa_structure_2008,lieberman2009comprehensive,halverson2014melt,nuebler2018chromatin,saintillan2018extensile,mahajan2022euchromatin,chan2024activity,shin2024transcription}, \emph{e.g.} proposing a mechanism for the stability of chromosome territories~\cite{rosa_structure_2008} and for coherent chromatin motion~\cite{saintillan2018extensile,mahajan2022euchromatin} (Fig.~\ref{fig:chromatin}b). In these models, the genetic material is generally represented as a bead-spring polymer, whose \textbf{persistence length}%
\begin{marginnote}[]
\entry{Persistence length}{Length scale above which a polymer or filament bends due to thermal fluctuations.}
\end{marginnote}%
is chosen to match either that of DNA or that of the chromatin fiber, depending on the scale of coarse-graining~\cite{halverson2014melt}. Some models also include the possibility of having different epigenetic states by representing chromatin as a \textbf{block copolymer},%
\begin{marginnote}[]
\entry{Block copolymer}{Polymer consisting of two or more distinct blocks (usually covalently linked) with specific characteristics.}
\end{marginnote}%
with different epigenetic domains represented by changing the monomer state and their interactions~\cite{jost2014modeling,michieletto2016polymer,jost2018epigenomics,shi2018interphase,michieletto2019nonequilibrium,shin2024transcription} (Fig.~\ref{fig:chromatin}c).

To simulate the action of type-II topoisomerase, an enzyme which consumes energy to resolve DNA entanglements and knots, the polymer strands can be allowed to cross each other with some energy penalty~\cite{naumova_organization_2013,goloborodko2016compaction,orlandini_synergy_2019,thornburg2026bringing}. The action of other molecules such as cohesin, condensin, or RNA polymerases can be included either implicitly, with active forces acting on polymer segments~\cite{saintillan2018extensile,mahajan2022euchromatin,shin2024transcription} (Fig.~\ref{fig:chromatin}b), or by modeling the proteins explicitly~\cite{goloborodko2016chromosome,goloborodko2016compaction,orlandini_synergy_2019,ryu2021bridging,chan2024activity}. The nucleoplasm mediates hydrodynamic interactions between chromatin segments; these are often assumed to be screened due to nuclear crowding, but can become important in the presence of active forces. In Ref.~\citenum{saintillan2018extensile}, the authors included the effect of active forces exerted by chromatin on the nucleoplasm, and later extended the model to include hydrodynamic interactions between distinct chromatin segments~\cite{mahajan2022euchromatin}. 

Models of \textbf{bacterial DNA} are mostly very similar to those used for eukaryotes~\cite{jun2006entropy,harju2024loop,thornburg2026bringing}, with the main difference being that bacterial genome is usually much smaller and often has a ring (instead of linear) topology, which affects \emph{e.g.} the entropic repulsion between DNA segments. Indeed, ring polymers repel more strongly than linear ones, since their topology and excluded volume interactions prevent their concatenation~\cite{jun2006entropy}.

More detailed coarse-grained models than those described above, representing chromatin at \textbf{sub-nucleosome resolution} ---including histones and histone tails--- have been used to study chromatin phase separation and material properties~\cite{farr2021nucleosome,zhou2025multiscale}. These models typically do not encode base-specific interactions, which are necessary to model processes such as DNA hybridization, and are captured \emph{e.g.} by oxDNA~\cite{poppleton2023oxdna} or three-sites-per-nucleotide models~\cite{hinckley2013experimentally}.

\subsubsection{Continuum models} 

\textbf{Active two-fluid models} adopt a similar formalism  to that of active gel theory (see also Sec.~\ref{sec:cytoskeleton_cm} and~\ref{sec:multicell_cm}). Within this approach, chromatin and the nucleoplasm can be modeled as two active and viscoelastic fluids~\cite{bruinsma2014chromatin,eshghi2023activity}, where the action of motors on the chromatin fiber is represented  by equal and opposite forces acting on chromatin and on the nucleoplasm/solvent (Fig.~\ref{fig:chromatin}d). \textbf{Field-theoretical models} represent chromatin domains with different properties, such as euchromatin and heterochromatin or different epigenetic states, as fields which evolve in time following a phenomenological free energy~\cite{lee2017new,laghmach2021interplay,michieletto2019nonequilibrium} or a master equation~\cite{olmeda2026scaling}. These models are able to qualitatively reproduce experimentally observed behaviors such as the euchromatin-heterochromatin microphase separation~\cite{laghmach2021interplay,michieletto2019nonequilibrium} (Fig.~\ref{fig:chromatin}e-f) and DNA methylation during embryonic development~\cite{olmeda2026scaling}. Some of these models take inspiration from classic statistical mechanics ones, such as the Ising model. In Ref.~\citenum{michieletto2019nonequilibrium}, for example, epigenetic marks are modeled as magnetic spins that tend to align with each other (Fig.~\ref{fig:chromatin}f). 

\subsubsection{Comparison}

The main advantage of particle-based models compared to continuum ones is that they make it possible to deal with the intrinsically multi-scale nature of chromatin. It is challenging to include in a single continuum model all the relevant length scales, such as the fiber diameter, its persistence length, but also higher-order structures such as loops and topologically associating domains~\cite{jost2018epigenomics}. Moreover, excluded volume interactions introduce complex correlations in three dimensions (such as entanglements) that are notoriously difficult to account for analytically. 

On the other hand, particle-based models suffer from size and time-scale limitations due to the extremely slow relaxation of long (and especially entangled) polymers~\cite{rosa_structure_2008,shi2018interphase,hildebrand2024mitotic}. Continuum models instead allow us to describe chromatin dynamics and organization on the scale of the whole nucleus, capturing features such as large euchromatin and heterochromatin domains, at the cost of a loss of resolution on intermediate-scale features. 

\subsubsection{Case study: Compaction of mitotic chromosomes by loop extruders} 
\label{sec:chromatin_cs}

In eukaryotes, interphase chromatin at the megabase scale is often modeled as a \emph{fractal globule}, where segments form compact, crumpled structures~\cite{lieberman2009comprehensive,mirny2011fractal}. After replication, the chromatin conformation changes dramatically, as the chromosome copies (\emph{sister chromatids}) are compacted into cylindrical mitotic chromosomes. Since the 1980s, experimental data had suggested that this compact structure was achieved \emph{via} the formation of chromatin loops, but the mechanism by which these were formed remained unclear~\cite{goloborodko2016compaction}.

In Ref.~\citenum{alipour2012self} it was shown with stochastic simulations of a simple 1D lattice model that chromatin loops can be stabilized by condensins, energy-consuming loop extruders. Although single condensins undergo constant binding and unbinding, multiple stacked ones can stabilize the loops. However, due to the limited system size, this model could not recapitulate the large-scale loop organization observed in experiments. Subsequent work combining a 1D continuum model and stochastic simulations showed that loop extruders alone can generate arrays of non-overlapping loops with a size distribution matching experiments~\cite{goloborodko2016chromosome}.

The remaining question was whether this mechanism could lead to the compaction and segregation of sister chromatids in 3D space observed in experiments. This problem is analytically intractable, due to the many relevant length scales involved and to the complex correlations introduced by excluded-volume interaction. To tackle this difficult challenge, Ref.~\citenum{goloborodko2016compaction} employed coarse-grained molecular dynamics simulations (Fig.~\ref{fig:chromatin}a). These showed that the combined action of loop extruders and topoisomerase-II (allowing strand crossing to resolve entanglements) was sufficient to lead to chromosome compaction and segregation in three dimensions~\cite{goloborodko2016compaction}. 

These studies show how simple continuum models can give significant insight into chromatin organization, but need to be complemented by particle-based simulations to capture the 3D topology and excluded-volume–induced correlations that govern chromatid organization.

\subsection{Biomolecular condensates}
\label{sec:condensates}

\begin{figure}
    \centering
    \includegraphics[width=\linewidth]{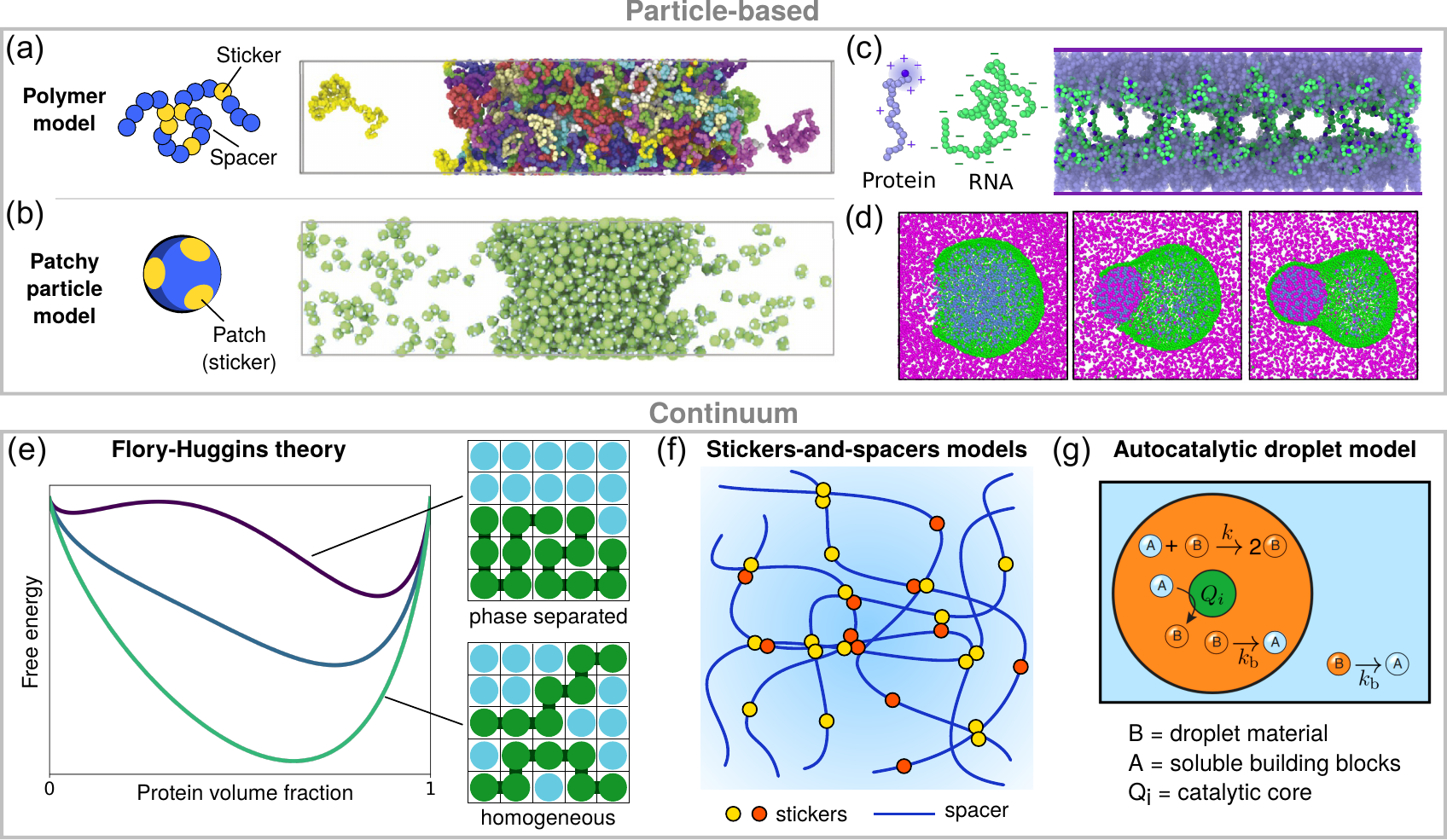}
    \caption{\textbf{Models for biomolecular condensates.} \textbf{(a-b)} Schematics of a polymer (a) and patchy particle (b) model of a protein, and phase coexistence simulations with slab geometry employing the two different models. Panels adapted with permission from Reference~\citenum{espinosa2020liquid}. \textbf{(c)} RNA (green, $-$~charge) forms bridges between two protein brushes on the surface of chromosomes (purple, $+$~charge), generating an attractive force between them. Simulations realized with the model of Reference~\citenum{sorichetti2025charge}. \textbf{(d)} Proteins (magenta) interacting with a vesicle (green) and endosomal lumen contained in the vesicle (blue). Panel adapted from Reference~\citenum{bussi2023stress} (CC BY 4.0). \textbf{(e)} Flory-Huggins free energy of a phase-separating polymer system and schematic representation of the corresponding lattice model (green: monomers, light blue: solvent). At low temperature or strong inter-monomer attraction, the system undergoes phase separation. \textbf{(f)} Schematic representation of a stickers-and-spacers model: flexible polymer chains (blue) are decorated with two types of attractive stickers (yellow and orange), which represent \emph{e.g.} different amino-acid residues. \textbf{(g)} Schematic of an autocatalytic droplet where droplet material B (orange) is produced from soluble building blocks A (light blue) \emph{via} the reaction $\mathrm{A}+\mathrm{B}\rightarrow2\mathrm{B}$ and also the reaction $\mathrm{A}\rightarrow\mathrm{B}$ at the catalytic core (green). Panel adapted with permission from Reference~\citenum{zwicker2015suppression}; copyright 2015 American Physical Society.}
    \label{fig:cond}
\end{figure}

Biomolecular condensates, or membraneless organelles, are  viscoelastic liquid droplets resulting from the liquid-liquid phase separation of proteins and nucleic acids~\cite{Brangwynne2015,shin2017liquid}. Condensates perform key cellular functions, spatially organizing chemical reactions, regulating  gene expression, and controlling stress response~\cite{shin2017liquid}. The proteins involved generally alternate short groups that significantly attract each other, contributing to phase separation (\emph{stickers}), and longer ones that mainly interact \emph{via} volume exclusion (\emph{spacers})~\cite{choi2020physical} (Figs.~\ref{fig:cond}a and f). The fundamental question is how the biochemistry of the individual interacting macromolecules determines the physical properties of the resulting assembly. Models retaining the discreteness of proteins and nucleic acids are ideal to bridge between the nanoscale and the microscale, which is in turn more easily understood in terms of continuum theories.

\subsubsection{Particle-based models} 
\label{sec:condensates_pb}
As the proteins involved in the formation of biomolecular condensates often have large disordered regions, they can be represented as flexible bead-spring polymers (Fig.~\ref{fig:cond}a). \textbf{Residue-resolution models}~\cite{maristany2025modeling}, such as the HPS model~\cite{dignon2018sequence,das2020comparative}, Mpipi~\cite{joseph2021physics} and CALVADOS~\cite{tesei2021accurate} model proteins in a top-down way, representing each amino-acid residue with a single bead. The interaction parameters are usually optimized to quantitatively match the experimentally measured values of certain properties, such as the protein radius of gyration~\cite{dignon2018sequence,tesei2021accurate}, or by employing bottom-up methods based on all-atom simulations or bioinformatics data~\cite{joseph2021physics}. Residue-resolution models have achieved near-experimental accuracy in predicting protein phase diagrams~\cite{das2020comparative,tesei2021accurate,joseph2021physics}. Models with \textbf{resolution coarser than the single residue} are used when high chemical specificity is unnecessary, or when there is a practical need to simulate larger systems~\cite{choi2019lassi,Ranganathan2020,Ranganathan2022,hernandez2024liquid,sorichetti2025charge} (Fig.~\ref{fig:cond}c). Space can be discretized to make the simulations more efficient, making polymers live on a 2D~\cite{freeman2017eukaryotic,Ranganathan2020} or 3D lattice~\cite{feric2016coexisting,Harmon2017,choi2019lassi,martin2020valence}. Lattice-based Monte Carlo simulations employing the LASSI~\cite{choi2019lassi} and PIMMS models~\cite{martin2020valence} have helped understand the phase behavior of multi-component condensates~\cite{feric2016coexisting}, and clarified the role of patterning and valence of aromatic residues~\cite{martin2020valence}.

A higher level of coarse-graining is achieved by representing the proteins as single beads with discrete interaction sites (\textbf{patchy particles})~\cite{freeman2017eukaryotic,ghosh2019three,espinosa2020liquid,joseph2021thermodynamics,zeng2021plc,sun2022kinetic,berthin2025microscopic} (Fig.~\ref{fig:cond}b). This simplifies steric repulsion and eliminates entanglements, allowing the simulation of systems containing thousands of proteins over larger time scales. While Monte Carlo simulations can connect molecular valences and affinities to thermodynamic phase diagrams~\cite{ghosh2019three}, molecular dynamics can be used to address the role of kinetics, and has helped understand condensate size, stability and surface tension~\cite{espinosa2020liquid,zeng2021plc,sun2022kinetic}. The simplicity of single-particle models makes them also ideal to study condensates interacting with other biological structures, such as chromosomes~\cite{hernandez2024liquid,sorichetti2025charge} (Fig.~\ref{fig:cond}c) and membranes~\cite{satarifard2018nanodroplets,bussi2023stress} (Fig.~\ref{fig:cond}d). 

\subsubsection{Continuum models} 
\label{sec:condensates_cm}
The most successful theoretical framework to describe the thermodynamics of condensates is the \textbf{Flory-Huggins theory} of polymer mixtures~\cite{Brangwynne2015}. Within this framework, starting from a lattice model (where each site represents a monomer or a solvent molecule), one can obtain a mean-field expression for the free energy of mixing, from which one can derive the phase diagram of the system if the Flory parameter can be estimated~\cite{lee2013spatial} (Fig.~\ref{fig:cond}e).%
\begin{marginnote}[]\entry{Flory parameter}{Mean-field parameter quantifying the balance between polymer-polymer and polymer-solvent interactions. A large Flory parameter favors phase separation.}\end{marginnote}%
Classic Flory-Huggins theory, however, does not capture effects such as localized stickers, which require \textbf{stickers-and-spacers models}~\cite{wang2018molecular,choi2020physical,choi2020generalized} (Fig.~\ref{fig:cond}f). These are based on the mean-field theory of associative polymers, \emph{i.e.} polymers with attractive groups~\cite{semenov1998thermoreversible}. Other approaches to modeling localized stickers include the \textbf{Statistical Associating Fluid Theory (SAFT)}, describing the thermodynamics of particles with attractive patches~\cite{Jacobs2014,sanders2020competing}. Identifying the free energy minima is a hard task in mixtures with a large number of components, and is often done numerically. Studies have shown that entropy often favors homogeneous states over phase-separated ones in multicomponent mixtures~\cite{Jacobs2017,Mao2019}. The time evolution of condensate interfaces can be derived from the free energy using the \textbf{Cahn-Hilliard formalism}~\cite{cahn1958free}, which has been also employed to study how condensates exert forces on chromatin~\cite{strom2024condensate} and cytoskeletal filaments~\cite{boddeker2022nonspecific}. \begin{marginnote}[]\entry{Cahn-Hilliard model}{Mean-field model describing phase separation, which combines bulk and interface free energies to model domain evolution.}\end{marginnote}

Condensate stability is influenced by active processes such as energy-consuming \textbf{chemical reactions}. These active processes have been shown to be an important mechanism to suppress coarsening and lead to the stable size-limited condensate droplets observed in cells~\cite{zwicker2014centrosomes,zwicker2015suppression,wurtz2018chemical,soding2020mechanisms,kirschbaum2021controlling,zwicker2022intertwined}. In turn, condensates can tune reaction rates by locally increasing the concentration of chemical species~\cite{shin2017liquid}. This complex interplay has been studied with continuum models where fluxes of chemically reactive molecules within multicomponent condensates are described using the reaction-diffusion formalism~\cite{zwicker2014centrosomes,zwicker2015suppression,wurtz2018chemical,soding2020mechanisms,kirschbaum2021controlling,zwicker2022intertwined}. This framework has been applied for instance to understand the growth of centrosomes, which have been shown to behave as autocatalytic droplets controlled by enzymatic activity~\cite{zwicker2014centrosomes,zwicker2015suppression}(Fig.~\ref{fig:cond}g).

Continuum models have also been instrumental to study how condensates interact with cellular biopolymers such as chromatin and the cytoskeleton. In these models, the growth of condensate droplets is limited by the surrounding viscoelastic polymer medium~\cite{shin2018liquid,rosowski2020elastic}. Under some conditions, this has been predicted to lead to size selection and equilibrium microphase separation~\cite{ronceray2022liquid,qiang2024nonlocal,fernandez2026thermodynamics}. 

\subsubsection{Comparison} Despite their usefulness in predicting the qualitative phase behavior of biomolecular condensates, mean-field continuum models have several limitations. Flory-Huggins theory is best suited to homopolymers, rather than proteins with localized sticky sites. Continuum stickers-and-spacers models are better suited for these systems, but they usually treat spacers as phantom chains with no excluded volume, including the latter only in an effective manner. Thus, chain connectivity and sequence patterning (the microscopic arrangement of stickers) are not captured directly~\cite{choi2020physical}. 

Additionally, it is challenging to deal with systems with a large number of different protein species~\cite{choi2020physical}, whereas in particle-based models the computational cost only depends on the total number of simulated particles rather than on the number of species. Particle-based models, on the other hand, can suffer from slow dynamics, especially in dense phases~\cite{biswas2024molecular}, which can make it hard to explore experimentally relevant system sizes and time scales. Moreover, chemical reactions and processes such as turnover are not straightforward to simulate, usually requiring a combination of molecular dynamics and Monte Carlo methods~\cite{berthin2025microscopic}, while they can be more easily included in a continuum formalism~\cite{kirschbaum2021controlling}.

\subsubsection{Case study: Fundamental interactions driving the phase separation of disordered proteins} 

A central question in the study of biomolecular condensates is whether general physicochemical principles can be identified governing their phase behavior: which interactions between amino-acid residues are the main drivers of phase separation, and how is the latter modulated by sequence composition and protein charge?

In this context, a well-studied class of model systems is that of RNA-binding proteins containing a disordered, low-complexity \emph{prion-like domain}, such as hnRNPA1 and the \emph{FUS family proteins}~\cite{wang2018molecular,bremer2022deciphering}. These proteins are key components of stress granules and other RNA--protein condensates, and defects in their phase separation have been associated with the onset of neurodegenerative diseases~\cite{patel2015liquid}. Two landmark studies addressed the question of which amino-acid interactions are the main drivers of phase separation in these proteins, combining experiments and continuum models~\cite{wang2018molecular,bremer2022deciphering}.

Employing a minimal stickers-and-spacers framework, Ref.~\citenum{wang2018molecular} identified the cation--$\pi$ interactions between arginine and aromatic residues (particularly tyrosine) as the main drivers of phase separation of FUS family proteins. Studying the prion-like domain of hnRNPA1, Ref.~\citenum{bremer2022deciphering} used a generalized stickers-and-spacers model accounting for cooperative effects and different sticker--sticker interaction strengths, and identified the $\pi$--$\pi$ interactions between aromatic residues as the main contributors to phase separation, with arginine being a context-dependent auxiliary sticker. This work also showed that deviations of the net charge per residue from near-electroneutrality weaken phase separation.

Particle-based simulations employing a residue-resolution model (Mpipi)~\cite{joseph2021physics} favored the view that arginine--aromatic interactions are stronger drivers of phase separation than aromatic--aromatic ones, in line with Ref.~\citenum{wang2018molecular}, while also reproducing the measured phase behavior of the hnRNPA1 variants of Ref.~\citenum{bremer2022deciphering}. This particle-based approach additionally suggested that the relative strengths of these interactions are context-specific: their precise ranking matters for the phase behavior of some proteins, such as FUS, but not of others, such as hnRNPA1.

Together, these studies show how experiments, continuum models, and particle-based simulations can be combined to resolve the relative importance of specific interactions between residues, and to quantify how sequence composition and charge modulate their effects on phase separation.

\subsection{Biological tissues}
\label{sec:multicell}

Biological tissues (\textbf{confluent} and non-confluent) are multicellular systems widely studied to understand development, self-organization, wound healing, and cancer. These energy-consuming systems consist of cells that contain all the subsystems discussed above. Biophysical models often simplify them by assigning cells effective properties such as contractility, motility, signaling, division, or apoptosis. \begin{marginnote}[]\entry{Confluency}{Degree to which units cover a given surface, with 100\% indicating a confluent state.}\end{marginnote}

\begin{figure}[t]
\includegraphics[width=\textwidth]{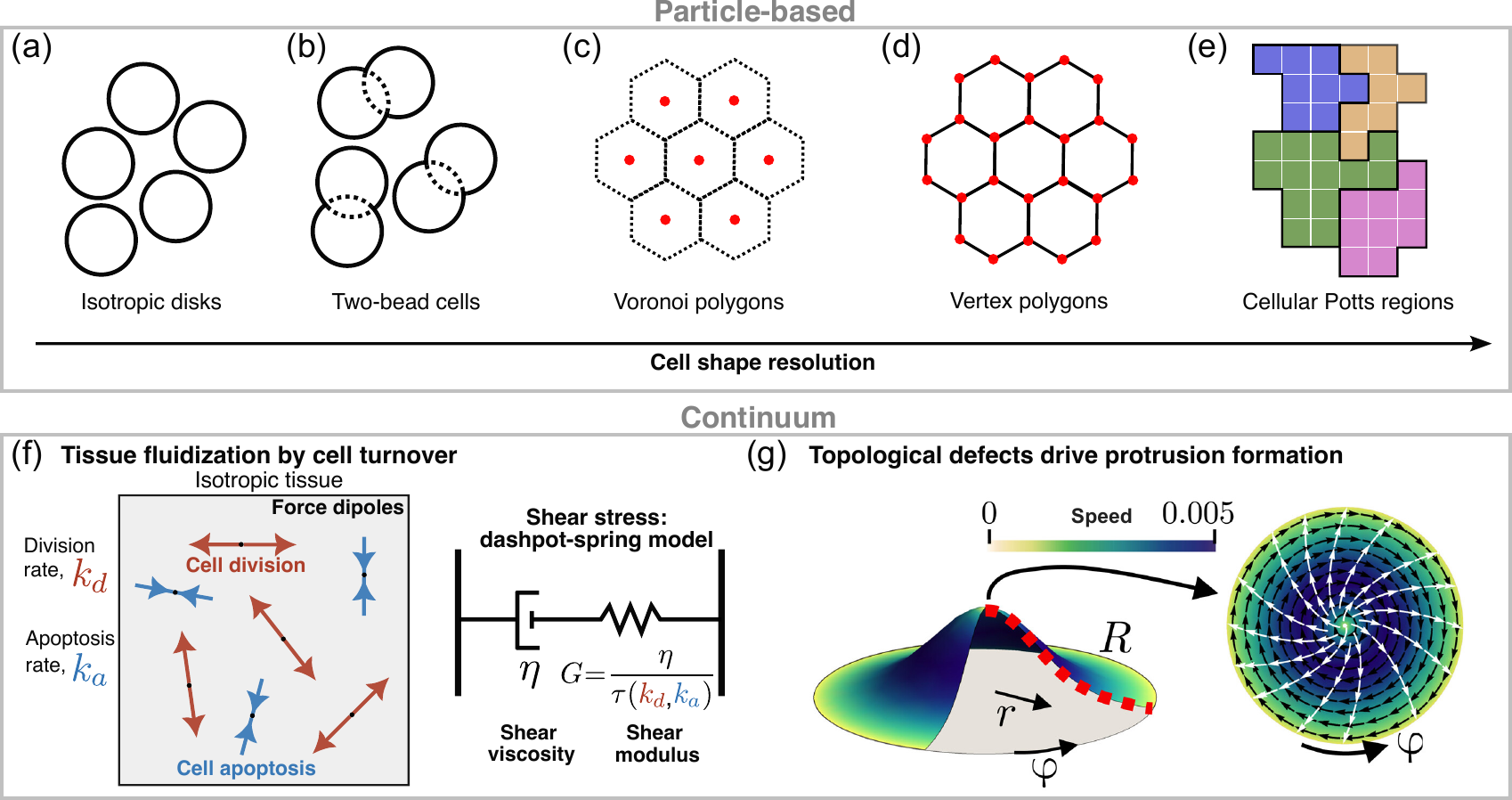}
\caption{\textbf{Models for biological tissues.}
\textbf{(a-e)}: Particle-based models with increasing cell shape resolution from left to right, where each cell is represented as a disk \textbf{(a)}, two joined beads \textbf{(b)}, a polygon emerging from tessellation of cell centers \textbf{(c)}, a polygon defined by all its vertices \textbf{(d)}, or a set of lattice sites \textbf{(e)}, respectively.
\textbf{(f)} Tissue under cell turnover, with division rate $k_d$ and apoptosis rate $k_a$. Modeling these events as force dipoles (red and sky-blue arrows), the authors of Reference~\citenum{ranft2010fluidization} showed that tissue shear stress follows a Maxwell model of viscoelasticity: a dashpot with a viscosity $\eta$ in series with a spring with a shear modulus $G$. The long-time fluid behavior is controlled by turnover rates ($k_d$ and $k_a$).
\textbf{(g)} A continuum active polar fluid model shows how +1 topological defects induce a planar-to-buckled instability of a clamped layer, with the instability amplified in the case of extensile activity and inhibited in the case of contractile activity. Panel reproduced with permission from Reference~\citenum{hoffmann2022theory} (Copyright 2022, Science Advances, AAAS). Left: sketch of the buckled layer with a +1 defect at the center. Right: top view. Black arrows represent the azimuthal velocity field, the color scale its magnitude, and the white arrows the director field.  This model provides a quantitative framework to understand protrusion growth in C2C12 myoblast monolayers~\cite{guillamat2022integer}.}
\label{fig:cellular}
\end{figure}

\subsubsection{Particle-based}
\label{sec:multicell_pb}

Particle-based models describe tissues as collections of interacting discrete particles. In the simplest model, cells are treated as \textbf{self-propelled polar particles} (Fig.~\ref{fig:cellular}a), aligning their velocity and polarity~\cite{szabo}, which helps explain cellular flocking and the role of specific cell-cell interactions in coordinated migration. However, these models are limited when cell proliferation or shape-dependent interactions become relevant.

Minimal 3D models of growing  aggregates use \textbf{two particles per cell}, with a repulsive force controlling the inter-particle distance (cell size) and division~\cite{Basan2013} (Fig.~\ref{fig:cellular}b). Despite their simplicity, they can capture experimentally-observed mechanical regulation of proliferation~\cite{montel2011stress} as well as the existence of a \textbf{homeostatic pressure}~\cite{basan2011dissipative}.%
\begin{marginnote}[]\entry{Homeostatic pressure}{Mechanical pressure a tissue builds up in a confined environment when cell birth and death are balanced.}\end{marginnote}%
They can easily accommodate free boundary conditions, reproducing  experiments on edge-dominated growing patterns~\cite{montel2011stress}, or wound healing~\cite{Basan2013}.

In confluent tissues, the shape of the cell conveys information about mechanical stresses between neighboring cells. Particle-based models use diverse strategies to capture different levels of cell shape resolution, as we explain below. 

A widely used class of frameworks for representing cell shape is \textbf{polygonal (or polyhedral) networks}, with Voronoi~\cite{honda1978description} (Fig.~\ref{fig:cellular}c) and vertex~\cite{Honda1980,Farhadifar2007} (Fig.~\ref{fig:cellular}d) models being the two most common ones. For higher-resolution cell shapes, the cellular Potts model~\cite{graner1992simulation,maree2007cellular} (Fig.~\ref{fig:cellular}e) can be employed. All of these frameworks simulate the time evolution of the tissue configuration assuming mechanisms such as adhesion and cortical tension, as well as area or volume constraints.

In \textbf{vertex models}, the positions of the vertices evolve continuously as a result of minimization of cell elastic energy and possible active forces, triggering cellular rearrangements when a cell-cell junction shrinks below a threshold. These models have revealed how mechanical forces drive embryonic morphogenesis (reviewed in~\cite{fletcher2014vertex}), as well as  processes such as wound healing~\cite{tetley2019tissue}. One of their classical predictions is that cell shape can predict tissue properties, such as solid-to-fluid transitions~\cite{bi2015density}, which was confirmed experimentally~\cite{park2015unjamming}. Mechanochemical feedbacks can be incorporated, where changes in the size of cells or their shared boundaries modulate the rate of change of elastic parameters over time. Such models can reproduce multicellular mechanical waves~\cite{boocock2023interplay, perez2024excitable}, predict active cellular rearrangements underlying tissue elongation~\cite{sknepnek2023generating}, and capture the formation of complex multicellular organizations in 2D~\cite{perez2023tension} and 3D~\cite{okuda2018combining}. An energy-minimization approach is implemented in Surface Evolver~\cite{brakke1992surface}, while time-dependent ODE solvers are implemented in Chaste~\cite{cooper2020chaste} and in custom solvers.
\begin{marginnote}[]\entry{Voronoi tessellation}{Partition of a space into adjacent non-overlapping polygonal regions based on the proximity to seed points.}\end{marginnote}

\textbf{Voronoi models} dynamically update only cell centers and rebuild the tissue network via \textbf{Voronoi tessellation}. This introduces two natural limitations. First, vertices are always shared by exactly three cells, not allowing the representation of rosettes ---configurations where more than three cells meet at a single point. Second, cell-neighbor exchanges cannot be explicitly controlled, despite being tightly regulated in biological tissues~\cite{curran2017myosin,brauns2024geometric,sknepnek2023generating}. Furthermore, the implicit determination of the cell boundaries limits their flexibility in handling complex boundary conditions compared to other cell-based models. Despite these limitations, Voronoi models are computationally efficient while still capable of reproducing realistic cell shapes~\cite{kaliman2016limits}, capturing packing, motility, glass-like transitions~\cite{Bi2016}, and the emergence of orientational order linked to cell division and motility~\cite{tang2024cell}.

The \textbf{cellular Potts model} was originally developed to study cell sorting~\cite{graner1992simulation}, and later extended to include processes such as cell growth, division, and motility~\cite{maree2007cellular}. Each cell is represented by a group of lattice sites. The canonical model includes energy terms penalizing changes in cell size and controlling adhesion, evolving tissue configurations via a modified Metropolis algorithm, as implemented in CompuCell3D~\cite{swat2012multi} or Morpheus~\cite{starruss2014morpheus}. At each step, lattice sites can change cell according to a Monte Carlo scheme, thus effectively redefining the contour of each cell. Its lattice-based Monte Carlo nature prevents precise geometrical and temporal control of cellular rearrangements and overall tissue dynamics. Additionally, the specific choice of the temperature parameter can lead to a number of non-biological artifacts such as fragmentation. Some of these limitations have been addressed in extensions of this model, by introducing connectivity constraints~\cite{durand2016efficient} and by reformulating the dynamics to introduce physically meaningful timescales~\cite{belousov2024poissonian}.

\subsubsection{Continuum models} 
\label{sec:multicell_cm}

Continuum models describe tissues through continuous fields. These fields can be used to represent individual cells at high shape resolution, as in phase-field models~\cite{nonomura2012study}, or to describe tissue-scale behavior without resolving individual cells.

\textbf{Phase-field models} of tissues typically assume that cellular dynamics can be separated into i) a fast relaxational mode associated with cell shape and ii) a slow translational mode associated with cell motion. Each cell is described by a smooth scalar field equal to 1 inside the cell, 0 outside the cell, and 1/2 at the boundary. These fields evolve over time following advection and relaxational dynamics with a free energy including single-cell and cell-cell terms~\cite{nonomura2012study,camley2017physical,balasubramaniam2021investigating}. These models have been particularly useful for studying collective cell motility~\cite{camley2017physical} and active nematic behavior in epithelial monolayers~\cite{balasubramaniam2021investigating}.

At the tissue scale, continuum models describe tissue through continuous fields, such as cell density, stress and polarization, and incorporate key biological processes like contractility, proliferation, and motility, guided by symmetries and conservation laws. For instance, Ref.~\citenum{shraiman2005mechanical} modeled tissue growth by treating local displacement, growth rate, and pressure as smooth fields, showing that a negative feedback from mechanical pressure onto growth could stabilize uniform growth ---a result consistent with the roughly uniform growth observed in the \emph{Drosophila} wing pouch during late larval stages~\cite{mao2013differential}.

Another class of models is \textbf{active gel models}~\cite{Kruse2005,Prost2015}, originally developed to study the cytoskeletal cortex (Sec.~\ref{sec:cytoskeleton_cm}). In these models, tissues are described as active viscoelastic materials that exhibit solid- or fluid-like behavior depending on the timescale of the process under study. At short timescales, or when cell-cell junctions maintain a persistent mechanical reference state, tissues respond as active elastic solids; at sufficiently long timescales, cell division, apoptosis, and rearrangements progressively relax the mechanical reference state, fluidizing the tissue. The authors of Ref.~\citenum{ranft2010fluidization} formalized this idea by treating division and apoptosis events as force dipoles entering the stress dynamics as source terms, and showed that the tissue behaves as a Maxwell viscoelastic material, with cell turnover rates controlling the long-term fluid response (Fig.~\ref{fig:cellular}f). 

Modeling other phenomena requires the inclusion of \textbf{polar fields}. In epithelial spreading, cells migrate collectively in response to cues. In this context, the addition of a polarization field guiding the direction of active forces allows continuum models to capture experimentally observed dynamical patterns, such as swirling motion~\cite{kopf2013continuum} and fingering~\cite{kopf2013continuum,alert2019active}. Furthermore, the inclusion of mechanical feedback between strain and contractility within an active elastic solid description was shown to generate mechanical traveling waves during tissue spreading~\cite{banerjee2015propagating}. Reproducing more complex spatio-temporal patterns~\cite{bittig2008dynamics, ioratim2023mechanochemical} requires anisotropic extensions that incorporate polar and/or nematic stresses~\cite{salbreux2022theory,blanch2021quantifying}, taking ideas from the physics of liquid crystals. These anisotropic active gel models have also been used to study the role of topological defects, which are regions where the nematic order is disrupted, in promoting cellular extrusion~\cite{saw2017topological} and differentiation~\cite{guillamat2022integer}, and the formation and growth of swirling protrusions~\cite{guillamat2022integer,hoffmann2022theory} (Fig.~\ref{fig:cellular}g).

\subsubsection{Comparison}
The main strength of continuum frameworks lies in their simplicity. By reducing tissue dynamics to a few smooth fields subject to given symmetries, they can efficiently capture tissue-scale collective behavior, enhance intuition, and exchange ideas with other systems governed by similar symmetries ---such as liquid crystal physics, widely used to study active nematic tissues. However, they cannot resolve how individual cells contribute to these large-scale phenomena, and are inherently limited when local heterogeneities and geometric details play a significant role. This includes cases in which cell properties vary across space or change over time through shape-dependent mechanosensing feedback mechanisms, as well as when cells interact with complex boundary conditions such as obstacles, neighboring tissues, or wounds. In such cases, particle-based models can overcome these limitations, though at a higher computational cost. Fortunately, a range of particle-based approaches exists, allowing us to balance computational cost against the geometric detail required for a given question.

\subsubsection{Case study: Superelasticity of micropatterned epithelial domes}
Ref.~\citenum{latorre2018active} illustrates how different modeling approaches can complement each other in the context of biological tissues. Epithelial sheets forming curved domes exhibit superelasticity, characterized by a tensional plateau up to $\sim$300\% areal strain and by striking cell-level heterogeneity, where some cells barely deform while others reach strains of up to 1000\%. To explain these observations, the authors combined a series of modeling approaches. 

First, treating the curved dome as a continuum sheet under mechanical equilibrium, they used Laplace's law to extract surface tension from experimental measurements of dome radius and pressure. To understand the experimental tension-strain relation, they modeled the dome using a vertex model on a curved surface, using a Voronoi tessellation of experimentally measured cell centers as the initial configuration. While this particle-based approach reproduced the macroscopic tension-strain relation, it could not explain the cell-level shape heterogeneity. 

The idea for the missing mechanism came from active-gel models of cortex dynamics. Assuming conservation of the total cytoskeletal mass within each cell, and exploiting the fact that dome dynamics are much slower than actin turnover, the cortex can be taken at steady state, yielding a simple relation between cortical density and cell shape: as cells stretch, the cortex thins. This was incorporated into the vertex model as a modified constitutive relation in which tension is proportional to cortical density. The resulting model produced unbounded stretching, regularized by adding intermediate filaments as threshold-activated elastic springs, guided by their experimental observation in superstretched cells.

This work illustrates the power of integrating approaches: a continuum relation (Laplace's law) to characterize tissue-scale mechanics, a Voronoi tessellation to initialize a tissue configuration from minimal experimental data, a vertex model to explain the origin of the tension-strain relation, and a subcellular constitutive law rooted in active-gel theory to capture the mechanism driving cell-level heterogeneity.

\section{PARTICLE-BASED AND CONTINUUM MODELS: STRENGTHS AND LIMITATIONS}
\label{sec:comparison}

\begin{figure}
    \centering
    \includegraphics[width=\linewidth]{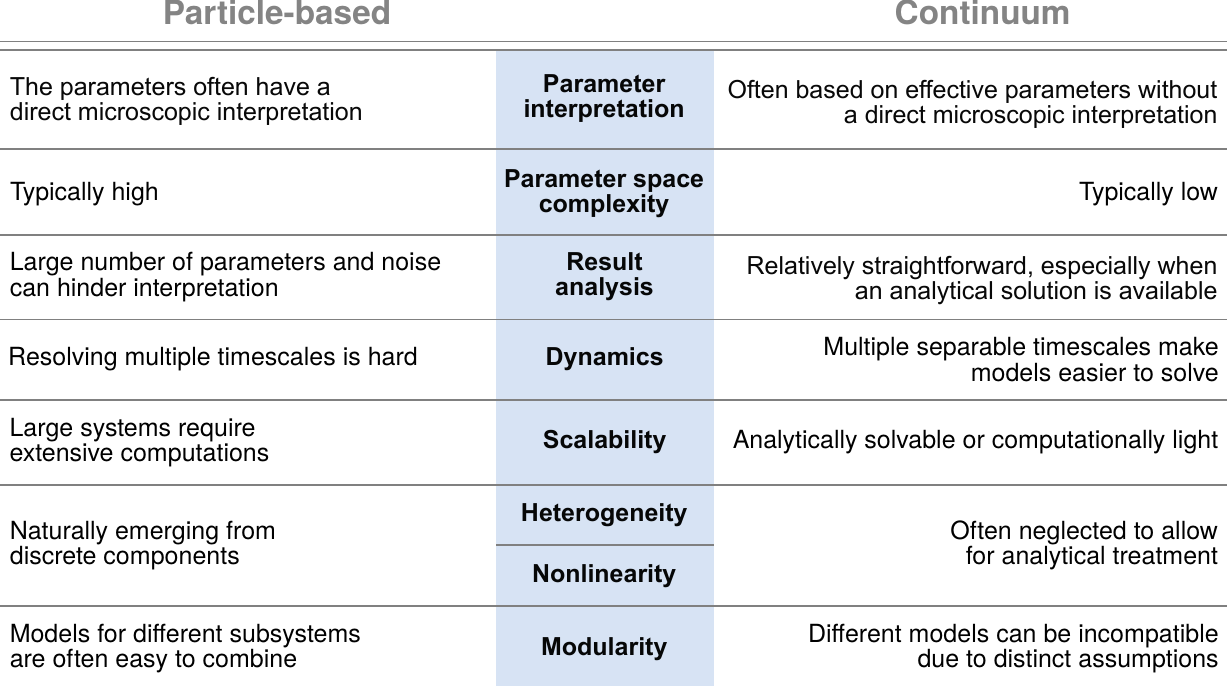}
    \caption{General comparison of particle-based and continuum models, highlighting the respective strengths and limitations.}
    \label{fig:comparison}
\end{figure}

Particle-based and continuum models offer complementary approaches to the modeling of biological systems, each with its own advantages and limitations. The choice between these two frameworks depends on the question of interest and on computational and analytical constraints, so that often there is not a unique best modeling choice. Below, we outline the key differences between these two approaches, which are also summarized in Fig.~\ref{fig:comparison}.

\begin{itemize}
\setlength{\itemsep}{2pt}      
\item \textbf{Parameter interpretation}: In continuum models, it can be difficult to relate effective parameters to microscopic mechanisms. For example, in a continuum model of the actomyosin cortex, contractility may be described by a single parameter~\cite{Prost2015,Reymann2016}, while in reality it depends on multiple factors such as myosin activity, actin density and local network structure. In a particle-based model, instead, individual molecules are resolved explicitly and contractility emerges from their mutual interactions. The parameter space is defined by quantities like rates of myosin binding and actin polymerization, of direct experimental interpretation~\cite{Nedelec2007, Descovich2018}.

\item \textbf{Parameter space complexity}: Many model parameters are not directly known from experiments, so exploring different parameter combinations is necessary to find the behavior that matches experimental observations. This tends to be easier in continuum models that rely on few phenomenological parameters. For example, in the Flory-Huggins model of protein phase separation (Sec.~\ref{sec:condensates_cm}), the mutual interactions between proteins are encoded by a single mean-field quantity, the Flory parameter. In particle-based models, interactions between all possible pairs of particles, representing different protein regions, must be set explicitly~\cite{joseph2021physics,tesei2021accurate}.

\item \textbf{Result analysis}: Because continuum models typically use fewer parameters and a higher level of coarse-graining, they often allow for a clearer interpretation of results. For instance, one may easily and analytically determine the phase boundaries for a continuum model, while for particle-based models specific simulation protocols or extensive data analysis might be necessary~\cite{maristany2025modeling}.

\item \textbf{Dynamics}: Particle-based models simulate the motion of individual particles, with time steps set by the fastest processes~\cite{frenkel2023understanding}. As a result, capturing both fast and slow dynamics can require long simulations. In continuum models, fast fields can be solved at fixed slow ones, yielding their dependence on the slow fields. This dependence is then substituted back in the slow field equations to eliminate the fast degrees of freedom~\cite{tailleur2022active,ranft2010fluidization}. This makes it possible to separate the timescales of the system, decoupling fast and slow processes.

\item \textbf{Scalability}: In particle-based models, the number of particles grows with the volume of the system, leading to increasing computational cost~\cite{allen2017computer,frenkel2023understanding}. Such cost can only be overcome by sacrificing details and coarsening the model. Continuum models are typically less demanding and, especially when analytically solvable, their complexity does not increase with system size.

\item \textbf{Heterogeneity}: 
Continuum models often rely on mean-field approximations, where interactions between biomolecules depend only on their average local environment~\cite{Lenz2020,choi2020physical}. This can result in neglecting correlation effects, which can be important in biological systems. For instance, coarse-graining biological polymers with a density field can completely miss the phenomenon of entanglement. Particle-based models, instead, naturally capture correlations and local heterogeneity~\cite{goloborodko2016compaction,Chugh2017,DarPalaia}.

\item \textbf{Nonlinearity}: Continuum models often assume linear couplings between variables, sometimes by design (as in active gel models, Sec.~\ref{sec:cytoskeleton_cm}), sometimes as an approximation to make complex equations analytically tractable (as in many Helfrich-derived models, Sec.~\ref{sec:membranes_cm}). As a result, they may fail to capture large excitations where the system responds nonlinearly. Particle-based models do not require such approximations and typically incorporate nonlinearity as an emergent property~\cite{Descovich2018,azadbakht2024nonadditivity}.

\item \textbf{Modularity}: Particle-based models describe individual components explicitly, which makes them naturally modular. Models of different subsystems, such as the cytoskeleton and the plasma membrane~\cite{Sciortino2025,Ni2021} or intracellular filaments and a tissue~\cite{latorre2018active}, can thus be combined, provided they share a similar level of coarse-graining. In contrast, continuum models sometimes rely on incompatible assumptions or solution methods, making their integration more challenging. For example, a membrane described as a bendable 2D surface may be difficult to couple consistently to a 3D model of the cytoskeleton.

\end{itemize}

\section{CONCLUSIONS AND OUTLOOK}

Discrete and continuum models represent established complementary approaches for quantitatively studying cell and tissue biophysics. The question of which one is best suited to a given experimental system admits no universal answer. We compared these two classes of models for specific systems ---from the cytoskeleton to tissues--- discussing their capabilities and limitations, showing that the optimal choice depends on the research question, problem scales, and available data. Future advances, possibly addressing the challenges and perspectives discussed below, will come from both mastering and boosting each approach and from strategically coupling them.

\begin{summary}[Current challenges]
\begin{enumerate}
\setlength{\itemsep}{6pt}      

\item Although specific theories have been developed to understand different types of active biological phenomena, a general \textbf{theory of non-equilibrium systems}, comparable in scope to equilibrium statistical mechanics, is currently missing. In particular, we lack principles that relate microscopic energy consumption to macroscopic observables such as transport and pattern formation. A promising avenue is that of \textbf{stochastic thermodynamics}~\cite{seifert_stochtherm_2012}, which provides quantitative relations between energy dissipation, irreversibility, and fluctuations, thus providing a way to connect non-equilibrium effects to biological function~\cite{cao_stochthermbio_2025}.

\item Cellular metabolism and signaling involve a vast network of \textbf{chemical reactions}, yet a general framework to incorporate them into particle-based models is still missing. Current coarse-grained approaches can only represent a limited subset of reactions, relying on simplified rules~\cite{gissinger2020reacter}. Extending these models to include more general reaction schemes would improve our understanding of cellular reaction networks.

\item Current biophysical models are often tailored to specific experimental systems, leading to a landscape of heterogeneous and often incompatible frameworks. This fragmentation hinders the development of unified, predictive models of cellular behavior. Establishing general principles to \textbf{systematically couple distinct modeling approaches} would provide a critical step toward multiscale, consistent descriptions of living systems.

\end{enumerate}
\end{summary}

\begin{issues}[Future perspectives]
\begin{enumerate}
\setlength{\itemsep}{6pt} 

\item In recent years, \textbf{Machine Learning} (ML) and \textbf{Artificial Intelligence} (AI) have introduced new tools and frameworks in biophysical modeling. One prominent application is the \textbf{development of coarse-grained potentials} from atomistic models~\cite{noe2020machine,majewski2023machine}, which provides a new bottom-up avenue for building particle-based models. This approach could potentially speed up model development, possibly in direct iteration with experiments.

\item When it comes to aiding the development of continuum models, ML and statistical inference methods can be employed to \textbf{learn the equations governing a biological system} by analyzing experimental or simulated data~\cite{frishman2020learning,bruckner2020inferring,supekar2023learning,schmitt2024machine,straube2026shape}.

\item ML and AI are a powerful tool to \textbf{sample complex statistical distributions}, potentially much more efficiently than what is achieved by molecular dynamics and Monte Carlo methods. This can be achieved through \textbf{ML-assisted Monte Carlo}~\cite{gabrie2022adaptive,delbono2025performance}, by using generative AI methods to \textbf{produce statistically representative trajectories}~\cite{jing2024generative,murtada2025md}, or to sample directly the underlying probability distribution~\cite{noe2020machine,olsson2026generative}. These methods still require training data from molecular dynamics and Monte Carlo simulations.

\item Combining continuum reaction–diffusion descriptions with particle-based models could advance our understanding of how cellular mechanics couples to signaling networks. Developing a general framework to \textbf{integrate reaction-diffusion formalisms with particle-based simulations} would enable efficient modeling of this mechanochemical coupling, for example in biomolecular condensates, DNA~\cite{thornburg2026bringing}, and the cytoskeleton~\cite{Popov2016}.

\end{enumerate}
\end{issues}

\section{DISCLOSURE STATEMENT}

The authors are not aware of any affiliations, memberships, funding, or financial holdings that might be perceived as affecting the objectivity of this review.

\section{ACKNOWLEDGEMENTS}

The authors thank Yongick Cho, Ferdinand Horvath and Fabrizio Olmeda for providing feedback on the manuscript, and Claudia Flandoli for help with creating Fig.~\ref{fig:part1_summary} based on an original concept by Valerio Sorichetti. This work was supported by the European Union’s Horizon 2020 research and innovation program (ERC grant agreement No.~802960 to A.\v{S}.; Marie Skłodowska-Curie grant agreement No.~101034413 to I.P.),  the European Union’s Horizon Europe research and innovation program (Marie Skłodowska-Curie grant agreement No.~101278760 to C.V.-C.), the Vallee Scholarship (to A.\v{S}.), the EMBO Young Investigator Programme (to A.Š.), and the NOMIS Foundation (to F.P.-V.).


\bibliographystyle{ar-style4}
\bibliography{library}
\end{document}